\documentclass[aps,pra,twocolumn,groupedaddress,showpacs,10pt]{revtex4-1}
\usepackage{braket}
\usepackage{graphicx}
\usepackage{dcolumn}
\usepackage{amsmath}
\usepackage{epstopdf}

\begin{document}

\title{Simulations of the bichromatic force in multilevel systems}

\author{L. Aldridge}
\author{S.E. Galica}
\author{E. E. Eyler}

\affiliation{Physics Department, University of Connecticut, Storrs, CT 06269}

\date{\today}

\begin{abstract}
Coherent optical bichromatic forces have been shown to be effective tools for rapidly slowing and cooling simple atomic systems. While previous estimates suggest that these forces may also be effective for rapidly decelerating molecules or complex atoms, a quantitative treatment for multilevel systems has been lacking.  We describe detailed numerical modeling of bichromatic forces by direct numerical solution for the time-dependent density matrix in the rotating-wave approximation.  We describe both the general phenomenology of an arbitrary few-level system and the specific requirements for slowing and cooling on a many-level transition in calcium monofluoride (CaF), one of the molecules of greatest current experimental interest.  We show that it should be possible to decelerate a cryogenic buffer-gas-cooled beam of CaF nearly to rest without a repumping laser and within a longitudinal distance of about 1 cm.  We also compare a full 16-level simulation for the CaF $B \leftrightarrow X$ system with a simplified numerical model and with a semiquantitative estimate based on 2-level systems.  The simplified model performs nearly as well as the complete version, whereas the 2-level model is useful for making order-of-magnitude estimates, but nothing more.
\end{abstract}


\maketitle

\section{Introduction}\label{sec:intro}

During the past few years, methods for direct laser slowing and cooling of small molecules have progressed remarkably, advancing from a demonstration of modest forces in 2009 all the way to a full realization of magneto-optical trapping at ultracold temperatures \cite{Shuman09,Barry12,Hunter12,Barry14,Hummon13,Yeo15,Tarbutt13,Zhelyazkova14}. Nevertheless, all-optical slowing and cooling is presently applicable only to a small number of molecules,  and the present schemes require complex optical configurations with multiple repumping beams to recycle atoms that are lost by unwanted radiative decay into distant ``dark" states.  In 2011 we proposed that the coherent optical bichromatic force could improve greatly on this situation, by providing not only a much stronger force, but one that allows much greater momentum transfer prior to radiative loss into non-cycling dark states \cite{Chieda11}.  Our 2011 paper relied largely on semi-quantitative arguments based on a two-level model, admittedly not fully adequate to describe the multilevel manifolds inevitably encountered in real molecules.  Here we verify and greatly extend this treatment by describing detailed numerical modeling of the bichromatic force (BCF) in realistic multilevel systems.

The bichromatic force in two-level systems has successfully been demonstrated in several atoms and is amply described in prior publications \cite{Grimm90,Soding97,Williams99,Williams00,Partlow04,Chieda11,Chieda12,ChiedaThesis,Liebisch12,Galica13}.  In brief, the BCF is the coherent force produced by a balanced counterpropagating pair of two-color cw laser beams.  Each two-color beam can be regarded as a beat note train produced by beams with frequencies $\omega \pm \delta$, for a carrier frequency $\omega$ centered close to the two-level resonant frequency.  If the laser power is adjusted so that each beat is approximately a $\pi$-pulse, and if the counterpropagating beat notes are sequenced properly, a large net force is produced by alternating cycles of coherent excitation and stimulated emission.  Larger bichromatic detunings $\delta$ produce a faster beat note period and thus a larger force, at the cost of a laser power requirement that scales as $\delta^2$.  In practice detunings of 100-400 times the radiative decay lifetime are workable, producing average forces larger than the radiative force by a similar multiple.  The velocity range of the force is also superior to the radiative force, extending over a range of $\delta/k,$ where $k$ is the wave vector for the transition of interest.  Recent developments in diode laser technology and rf electronics have made it relatively easy to produce the multi-frequency beams required for the BCF.

In a molecule the maximum attainable velocity reduction will usually be limited not by the velocity range of the force, but instead by unwanted radiative decay to non-cycling dark states, typically vibrational or rotational levels that are addressed neither by the BCF laser nor by added repumping lasers~\cite{Chieda11}.  Thus there is a dual motivation for using stimulated optical forces when working with molecules: to enhance the average force and more importantly, to increase the number of optical cycles that occur before the molecule is lost into an inaccessible vibrational or rotational level.

\begin{figure}
\includegraphics[width=0.9\linewidth]{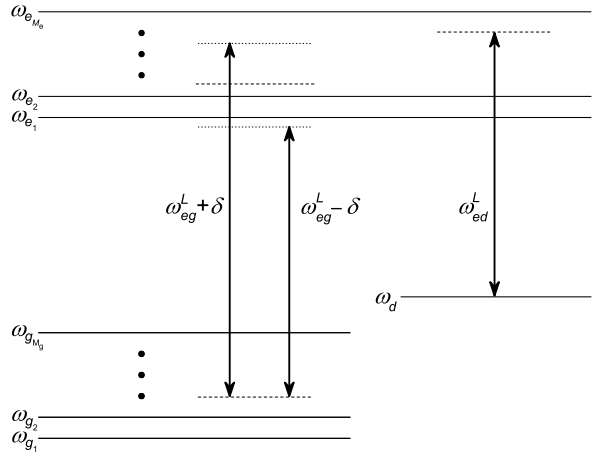}
\caption{A schematic diagram of the general multilevel system simulated here. Ground ($g$) and excited ($e$) levels with multiplicities $M_g$ and $M_e$ are coupled by a bichromatic laser field with frequencies $\omega^L_{eg}\pm \delta$, and excited levels are coupled to a distant ($d$) level by a laser field with frequency $\omega^L_{ed}$. There is no direct coupling or decay between the $d$ and $g$ states. Levels are labeled by their characteristic frequencies.}
\label{fig:multilevel}
\end{figure}

In a multi-level system like the one depicted in Fig. \ref{fig:multilevel}, it is impossible to achieve a $\pi$ pulse for every component of a transition, even for a stationary molecule with an optimal laser pulse.  Thus the optimal laser power and detuning involve unavoidable compromises, and the force is an average over the effects of several concurrently cyling transitions, some of them coupled coherently.  In our 2011 paper we treated this problem phenomenologically for the CaF molecule \cite{Chieda11}, an approach repeated more recently for MgF by Dai and coworkers \cite{Dai15}.  By contrast, in this paper we describe detailed numerical calculations that fully incorporate multilevel coherence, optical pumping, and the effects of repumping lasers that can optionally be used to repopulate the cycling system from dark states.  We begin with some general predictions and observations for three-level and many-level systems, then focus specifically on the CaF molecule, for which experimental tests are currently underway in our laboratory.

In Section \ref{sec:theory} we describe the theoretical approach used to numerically solve for the time-dependent density matrix of a multilevel system in a pair of bichromatic laser fields in the rotating-wave approximation.  Direct numerical integration has proven to be surprisingly stable and rapidly convergent, and we have successfully treated systems with up to 16 levels.  We start by treating three-level lambda-type systems in Section \ref{sec:twoplusone}, then introduce the effects of dark-state repumping from distant dark states in Section \ref{sec:repumping}.  In Section \ref{sec:dark_destabilization} we consider the additional complication of intra-system dark states, which can arise in many-level systems with high ground-state multiplicity.  Finally, in Section \ref{sec:caf} we consider in detail the feasibility of bichromatic slowing for the CaF molecule, a species of considerable current interest for laser slowing and cooling \cite{Zhelyazkova14}.  We treat the CaF problem using both full 16-level simulations for the $B\leftrightarrow X$ $(0-0)$ $P_{11}(1.5)/^PQ_{12}(0.5)$ branch and a more approximate approach, in which the system is simplified by dividing it into several subsystems that are assumed to cycle nearly independently.  We show that the subsystem-based modeling yields results very similar to the exact treatment in most cases, supporting the use of similar simplifications for treating the bichromatic force in other molecular systems.  More generally, the results verify the predictions of the semi-quantitative reasoning in our 2011 paper indicating that BCF slowing of molecular beams shows great promise.

\section{Theory and Methods}\label{sec:theory}

In our modeling of multilevel systems, we assume illumination by a pair of counterpropagating two-frequency laser beams that provide equal electric field magnitudes in all four components. In each beam the two frequencies $\omega+\delta$ and $\omega-\delta$ produce a train of beat notes, as described above. The total on-axis electric field for this arrangement is approximately \cite{Galica13}
\begin{eqnarray}
\vec{E}_{\text{BCF}}(z,t) = & 4 & E_0 \,\text{Re}[\{\hat{\epsilon} ( \cos(k z) \cos (\delta t) \cos(\chi / 2)\nonumber\\ & + & i \sin(k z) \sin(\delta t) \sin(\chi /2) \}e^{-i \omega t}],
\label{eq:efield}
\end{eqnarray}
valid so long as the beat length $c/\delta$ is much longer than the length scale over which the system is evaluated.  In this approximation the phase difference $\chi$ between the electric fields of the counterpropagating beat notes remains approximately constant. $E_0$ is the amplitude of the electric field of each of the four beams and $\hat{\epsilon}$ is the polarization, assumed to be identical for all four components.

When this field is incident on an atom or molecule with transitions $\ket{j} \leftrightarrow \ket{i}$, each such transition is associated with a complex Rabi frequency defined by
\begin{equation}
\Omega_{ij}^{R}(z,t) \equiv \frac{\bra{i} \hat{d} \cdot \vec{E}(z,t) \ket{j}}{\hbar},
\label{eq:rabidef}
\end{equation}
where $\hat{d}$ is the electric dipole operator.  It is important to note that systems may have closely spaced transitions with differing dipole transition elements, so the same optical field will simultaneously drive multiple transitions at different Rabi frequencies. We separate each of these Rabi frequencies $\Omega_{ij}^{R}$ into two terms with amplitudes $\Omega_{ij}$ and $\Omega_{ij}^*$ that describe co-rotating and counter-rotating components in the rotating wave approximation:
\begin{equation}
\Omega_{ij}^{R}(z,t) = \frac{1}{2}(\Omega_{ij} e^{-i \omega t} + \Omega_{ij}^* e^{i \omega t}).
\label{eq:rotatingrabi}
\end{equation}

It is also often convenient to refer to the ``Rabi frequency amplitude" of a transition, which we define as
\begin{equation}
\Omega_{ij}^0 \equiv \frac{E_0}{\hbar}\bra{i} \hat{d} \cdot \hat{\epsilon} \ket{j}.
\label{eq:rabiamp}
\end{equation}

As shown in Fig.~\ref{fig:multilevel}, our simulations assume a system with a lower-state manifold of $M_g$ levels and an upper-state manifold of $M_e$ levels, for which each of the lower-state levels may be coupled to any of the upper-state levels by the BCF optical field, depending on selection rules and line strengths. A single distant dark state is at times included as well, where ``distant" in this setting means that couplings of this state to any of the others by the primary BCF optical field are negligible. These systems will be referred to in this work as $M_g+M_e$, with (or without) a distant dark state. Sums over a ground state index ($p$, $q$, $r$) should be understood to run from 1 to $M_g$, and sums over an excited state index ($i$, $j$, $k$) from 1 to $M_e$.

The general BCF Hamiltonian includes the zero-field energies $H_0$ and the electric dipole coupling due to the applied optical field. This Hamiltonian is given by
\begin{equation}
\label{eq:hamiltonian}
\frac{H}{\hbar} = \frac{H_0}{\hbar} + \left(\sum_{i,p}{\Omega_{e_i g_p}^R\ket{e_i}\bra{g_p}} + \sum_{i}{\Omega_{e_i d}^R\ket{e_i}\bra{d}} + \text{c.c.}\right),
\end{equation}
where
\begin{equation}
\label{eq:hamiltonionzero}
\frac{H_0}{\hbar}=\sum_i \omega_{e_i} \ket{e_i} \bra{e_i} + \sum_p \omega_{g_p} \ket{g_p} \bra{g_p} + \omega_d \ket{d}\bra{d}.
\end{equation}
Here $\ket{g_p}$ are the ground states, $\ket{e_i}$ are the excited states, and $\ket{d}$ is the distant state.

The time-dependence of the density matrix $\rho$ is computed by numerically solving the Liouville equations with dispersion in the rotating wave approximation. In particular, the density matrix and Hamiltonian are expressed in terms of the quantities \cite{Einwohner1976,PurvesThesis}
\begin{eqnarray}
\label{eq:rotatingframe}
\widetilde{\rho}_{e_i g_p}& \equiv &\rho_{e_i g_p} \, e^{i \omega_{eg}^L t} \nonumber \\
\widetilde{\rho}_{e_i d}& \equiv &\rho_{e_i d} \, e^{i \omega_{ed}^L t} \nonumber \\
\widetilde{\rho}_{d g_p}& \equiv &\rho_{d g_p} \, e^{i(\omega_{eg}^L - \omega_{ed}^L) t},
\end{eqnarray}
where $\omega_{eg}^L$ is the carrier frequency of the laser field coupling the excited and ground states and $\omega_{ed}^L$ is the carrier frequency of the laser field coupling the excited and distant states.

We now make the rotating wave approximation by neglecting all oscillatory terms in the equations of motion with frequencies given by a sum of optical frequencies. We also include total decay rates $\Gamma_i$ and specific channel decay rates $\gamma_{ip}$.  The resulting full set of equations of motion of the density matrix of an $M_g+M_e$ system with a distant state $\ket{d}$ is given by,

\begin{widetext}
\begin{eqnarray}
\dot{\rho}_{e_i e_i}& = &-\sum_{q} \text{Im}[\Omega_{e_i g_q}^*\, \widetilde{\rho}_{e_i g_q}] - \text{Im}[\Omega_{e_i d}^*\,
  \widetilde{\rho}_{e_i d}] - \Gamma_i \rho_{e_i e_i} \nonumber \\
\dot{\rho}_{e_i e_j}& = &\frac{i}{2}\left[2(\omega_{e_j} - \omega_{e_i})\rho_{e_i e_j} + \sum_{q} \left(\Omega_{e_j g_q}^*\,
  \widetilde{\rho}_{e_i g_q} - \Omega_{e_i g_q}\, \widetilde{\rho}_{e_j g_q}^{\,*}\right) + \Omega_{e_j d}^*\, \widetilde{\rho}_{e_i d} - \Omega_{e_i d}\, \widetilde{\rho}_{e_j d}^{\,*}\right] - \frac{\Gamma_i+\Gamma_j}{2}\rho_{e_i e_j} \nonumber\\
\dot{\widetilde{\rho}}_{e_i d}& = &\frac{i}{2}\left[2(\omega_{ed}^L + \omega_d - \omega_{e_i})\widetilde{\rho}_{e_i d} +
  \Omega_{e_i d} (\rho_{e_i e_i} - \rho_{d d}) - \sum_{q}\Omega_{e_i g_q}\, \rho_{d g_q}^* + \sum_{k \neq i}\Omega_{e_k d}\, \rho_{e_i e_k}\right] - \frac{\Gamma_i}{2}\widetilde{\rho}_{e_i d}\nonumber\\
\dot{\widetilde{\rho}}_{e_i g_p}& = &\frac{i}{2}\left[2(\omega_{eg}^L + \omega_{g_p} - \omega_{e_i})\widetilde{\rho}_{e_i g_p}
  + \Omega_{e_i g_p} (\rho_{e_i e_i} - \rho_{g_p g_p}) - \sum_{q \neq p}\Omega_{e_i g_q}\, \rho_{g_q g_p} + \sum_{k \neq i}\Omega_{e_k g_p}\, \rho_{e_i e_k} - \Omega_{e_i d}\, \rho_{d g_p}\right] - \frac{\Gamma_i}{2}\widetilde{\rho}_{e_i g_p}\nonumber\\
\dot{\rho}_{dd}& = &-\sum_k {\dot{\rho}_{e_k e_k}} - \sum_q {\dot{\rho}_{g_q g_q}}\nonumber\\
\dot{\widetilde{\rho}}_{d g_p}& = &\frac{i}{2}\left[2(\omega_{eg}^L + \omega_{g_p} - \omega_{ed}^L -
  \omega_{d})\widetilde{\rho}_{d g_p} + \sum_{k}( \Omega_{e_k g_p}\, \widetilde{\rho}_{e_k d}^{\,*} - \Omega_{e_k d}^*\, \widetilde{\rho}_{e_k g_p})\right] \nonumber \\
\dot{\rho}_{g_p g_p}& = &\sum_k (\text{Im}[\Omega_{e_k g_p}^*\, \widetilde{\rho}_{e_k g_p}] +
  \gamma_{kp}\, \rho_{e_k e_k})\nonumber\\
\dot{\rho}_{g_p g_r}& = &\frac{i}{2}\left[2(\omega_{g_r} - \omega_{g_p})\rho_{g_p g_r} +
   \sum_k (\Omega_{e_k g_r} \widetilde{\rho}_{e_k g_p}^{\,*} - \Omega_{e_k g_p}^*\, \widetilde{\rho}_{e_k g_r}) +
   \Omega_{d g_r}\, \widetilde{\rho}_{d g_p}^{\,*} - \Omega_{d g_p}^*\, \widetilde{\rho}_{d g_r}\right].
\end{eqnarray}
\end{widetext}

Defining $N=M_g+M_e$ and taking into account the Hermiticity and preserved trace of the density matrix, this is a system of equations in $N$ real and $(N^2 + N)/2$ complex independent variables, as any one of the $N+1$ diagonal density matrix elements can be described as a function of only the other $N$ states. Numerical solutions to this set of equations are computed using the built-in numerical differential equation solver of Wolfram Mathematica \cite{Mathematica}. For a given set of fixed parameters, the computation of the time-dependence of the density matrix for a sixteen-level system requires about thirty seconds on a moderately fast personal computer (Intel i7-3770 processor at 3.4 GHz) for a time range of 100/$\Gamma$, where $\Gamma$ is the smallest of the decay rates $\Gamma_i$.

The behavior of an atom or a molecule with velocity $v$ in the lab frame is modeled by making the replacement $z \rightarrow v t$ in the spatial dependence of each Rabi frequency. This assumes that only motion along the laser beam axis is relevant; effects of transverse velocity across an inhomogenous beam profile are not considered here.  As in previous work \cite{Chieda12,ChiedaThesis,Soding97} the force on the system at a particular velocity is then obtained by application of Ehrenfest's theorem which gives, for the Hamiltonian in Eq.~\ref{eq:hamiltonian},
\begin{equation}
\label{ehrenfest}
F = -\hbar\left( \sum_{i,p}{ \text{Re}\left[ \widetilde{\rho}_{e_i g_p}\nabla\Omega_{e_i g_p}^*\right] }+ \sum_{i}{\text{Re}\left[\widetilde{\rho}_{e_i d}\nabla\Omega_{e_i d}^*\right]}\right).
\end{equation}
Here the $\nabla \Omega_{ij}$ are obtained symbolically from Eqs. \ref{eq:efield}--\ref{eq:rabiamp} prior to the $z \rightarrow v t$ replacement discussed above. This expression gives an immediate general idea of how to achieve strong forces: strong coherences must be maintained and must change in phase with the driving field.

\section{2+1 `$\Lambda$' Systems} \label{sec:twoplusone}

\begin{figure}
\includegraphics[width=0.9\linewidth]{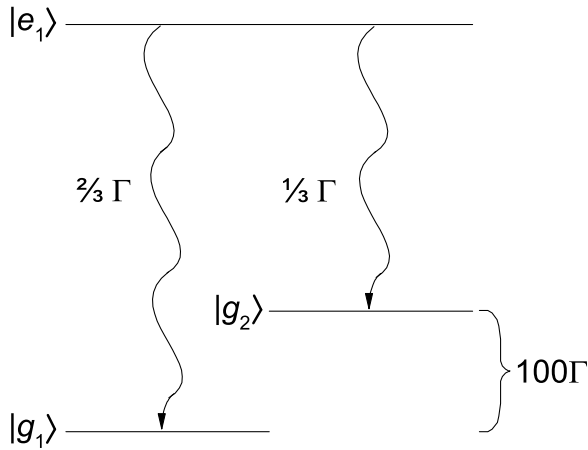}
\caption{The 2+1 system simulated in Section~\ref{sec:twoplusone}. Decay rates and ground-state splitting are indicated; the ratio of Rabi frequencies on each transition was set to the ratio of the square roots of the branching fractions.}
\label{fig:twoplusone}
\end{figure}

A simple 2+1 lambda-type system without a distant dark state was simulated to facilitate systematic study of the effects of multiple levels on the bichromatic force. The system is set up such that each of the two lower states can have a different electric dipole coupling to the excited state and a different zero-field energy (Fig.~\ref{fig:twoplusone}).

Since each transition has a different coupling to the electric field, the BCF laser irradiance produces two different Rabi amplitudes, $\Omega_{e_1g_1}^0$ and $\Omega_{e_1g_2}^0$. We assume in this section that each transition is optimally driven and that both laser fields have the same polarization, so that
 \begin{equation}
 \left(\frac{\Omega_{e_1g_1}^0}{\Omega_{e_1g_2}^0}\right)^2 = \frac{\gamma_{e_1g_1}}{\gamma_{e_1g_2}}.
 \end{equation}

 To aid comparisons, we define a total Rabi frequency amplitude $\Omega^{\textrm{tot}}$ as the quadrature sum of the two Rabi amplitudes,
\begin{equation}
\Omega^{\textrm{tot}} = \sqrt{\left(\Omega_{e_1g_1}^0\right)^2 + \left(\Omega_{e_1g_2}^0\right)^2}.
\end{equation}

There are two regimes in which the force resulting from BCF illumination is relatively easy to explain, the small-detuning regime ($\delta << |\omega_{g_1} - \omega_{g_2}|$) and the large-detuning regime ($\delta >> |\omega_{g_1} - \omega_{g_2}|$).

\begin{figure}
\includegraphics[width=0.9\linewidth]{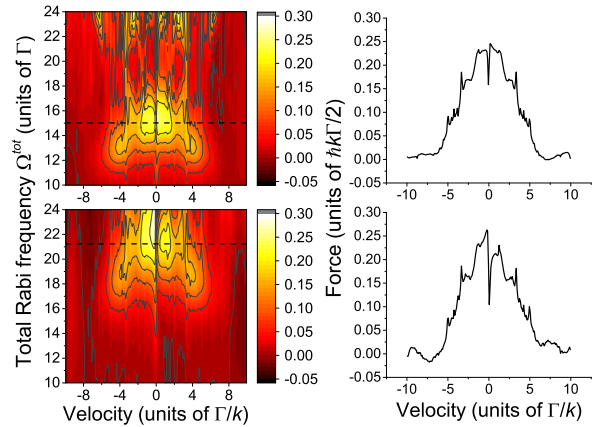}
\caption{(Color online) BCF force from simulations of a lamdba-type system, with the BCF carrier frequency resonant with the $\ket{g_1} \rightarrow \ket{e_1}$ transition (top) or the $\ket{g_2} \rightarrow \ket{e_1}$ transition (bottom). Force contours as a function of the total Rabi frequency $\Omega^\textrm{tot}$ and velocity (left) reveal the optimum $\Omega^\textrm{tot}$ for each transition (dashed line) which, though at different irradiances, result in similar force vs. velocity profiles (right).}
\label{fig:aiming}
\end{figure}

In the small-detuning regime, our simulations show that a central BCF frequency resonant with one transition or the other is preferable to one tuned near the weighted average energy of the lower states. When the carrier frequency $\omega$ is on resonance with one of the transitions, the other transition sees the light as essentially a single-frequency laser. Figure~\ref{fig:aiming} shows calculated force profiles for resonance with each of the ground-state levels $\ket{g_1}$ and $\ket{g_2}$.  While it is necessary to supply more laser irradiance on the weaker transition to achieve the optimal force at a given detuning, this force is comparable to the optimal force attainable on the stronger transition. This might be expected at first glance, given that the two-level BCF at optimal irradiance depends only on the bichromatic detuning. What may confound such an argument is the possibility of differing overall coupling strengths when the system reaches a periodic quasi-equilibrium.  Thus we discuss this equilibrium in some detail.

In a two-level system with resonant monochromatic illumination and without radiative decay, the population cycles between being fully in the ground state and fully in the excited state. When radiative decay is included, the system eventually damps to an equal mixture of the ground and excited states. With the addition of third state coupled to the excited state, assuming that the original transition is resonantly driven, the steady-state excited population becomes \cite{Sen15}
\begin{equation}
\Omega_{e_1g_1}^2 \Omega_{e_1g_2}^2 \Gamma \Delta_{e_1g_2}^2 / \mathcal{O}\left(\Omega_{e_1g_1}^6,\Omega_{e_1g_2}^4\right),
\end{equation}
where $\Delta_{e_1g_2}$ is the detuning of the light from the $\ket{e_1}\leftrightarrow\ket{g_2}$ transition. We see from this that for finite Rabi frequencies, if the second transition is driven with a non-zero detuning, the system remains at least partially bright (a full expression can be found in Ref.~\cite{Sen15}). Even though a true coherent dark state does not exist, the excited-state population is reduced compared to a resonantly-driven two-state system, and some population collects in the off-resonant state. For the system shown in Fig.~\ref{fig:twoplusone}, if both lower states were coupled to the excited state by the same monochromatic light  resonant with the $\ket{e_1} \leftrightarrow \ket{g_1}$ transition (and therefore $\Delta_{e_1g_2} = \omega_{g_2} - \omega_{g_1}$), the combined steady-state population in levels $\ket{e_1}$ and $\ket{g_1}$ would rise with increasing Rabi frequency from zero to a maximum of approximately two-thirds that occurs at a Rabi frequency beyond the small-detuning regime.  Likewise for the bichromatic case in the small-detuning regime, if the bichromatic carrier frequency is resonant with one transition, the other transition is expected to experience no appreciable bichromatic force. We expect in this case that the bichromatic force in the lamdba-type system will be reduced by a factor of the ``participating'' fraction, which can be estimated by adding the populations in the excited state and resonantly-coupled ground state under monochromatic illumination.

\begin{figure}
\includegraphics[width=1.0\linewidth]{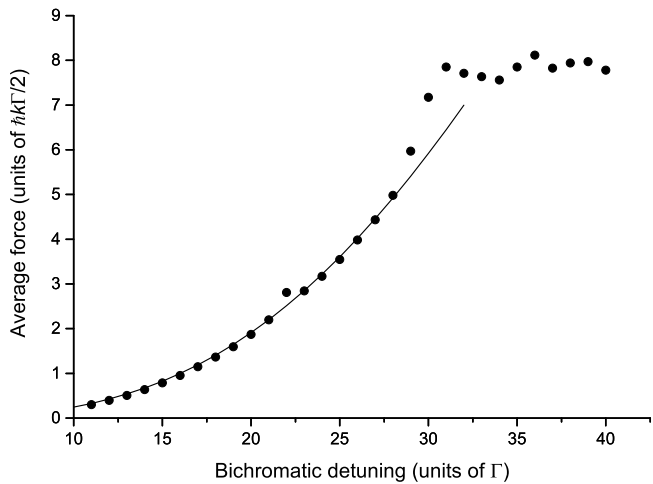}
\caption{Velocity-averaged force for BCF with carrier frequency resonant with the $\ket{g_1} \rightarrow \ket{e_1}$ transition of the system shown in Fig.~\ref{fig:twoplusone}. The Rabi frequency amplitude is $\Omega_{e_1g_1}^0=\sqrt{3/2}\,\delta$ throughout. Points show results of direct simulation and the line shows the two-level force times an estimated participating fraction calculated for monochromatic excitation (see text). Beyond $\delta\approx 25\Gamma$, the system transitions to the intermediate-detuning regime.}
\label{fig:smalldetuning}
\end{figure}

This is verified in Fig. \ref{fig:smalldetuning}, for which the example system is simulated using a range of small bichromatic detunings $\delta$, in each case keeping the carrier $\omega$ on resonance with the $\ket{g_1} \rightarrow \ket{e_1}$ transition and using the optimal Rabi frequency $\Omega_{e_1g_1}^0=\sqrt{3/2}\,\delta$.  In contrast to the two-level BCF, the average force is nonlinear in $\delta$, roughly scaling as $\delta^3$. Up to a bichromatic detuning of one-quarter of the ground-state energy splitting, this force closely follows a curve defined by the product of the two-level bichromatic force and the participating fraction calculated from the from the equilibrium populations for monochromatic excitation. An effective monochromatic Rabi frequency was included as a $\delta$-independent fitting parameter in calculating this participating fraction. The best-fit effective Rabi frequency was found to be 2.040(19) $\Omega^{\textrm{tot}}$.

The full-width at half-maximum (FWHM) of the force vs. velocity profiles was found to scale linearly with detuning, as it does for the two-level BCF, at detunings $\delta \leq 25\Gamma$. The best-fit FWHM is 0.706(12)$\,\delta/k$, somewhat wider at the half-maximum point than for two-level simulations \cite{Chieda12}, due primarily to a softening of the sharp velocity dependence near the edges of the force profile.

\begin{figure}
\includegraphics[width=1.0\linewidth]{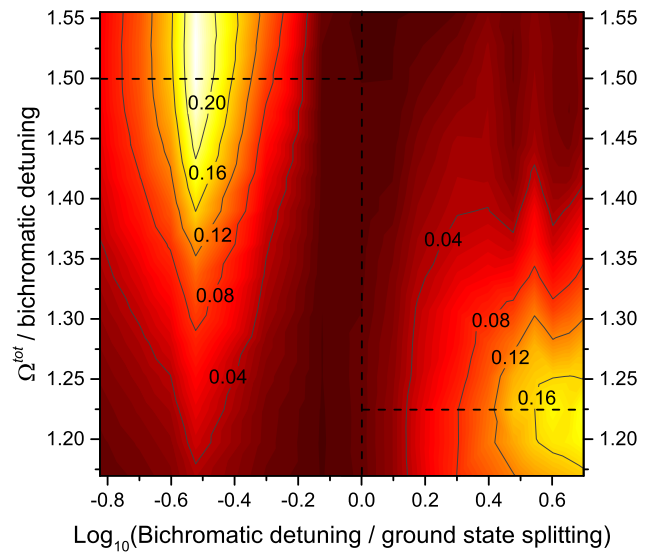}
\caption{(Color online) Velocity-averaged force in an asymmetric lambda-type system, scaled by the bichromatic detuning $\delta$, for a range of $\delta$. The contours are given in units of $\hbar k/2$. Results show a ``dead zone'' in the force near $\delta=|\omega_{g_1}-\omega_{g_2}|$ (vertical dashed line), and very different optimal values of $\Omega^{\textrm{tot}}/\delta$ in the small- and large-detuning regimes (horizontal dashed lines).}
\label{fig:intermediate}
\end{figure}

In the intermediate-detuning regime ($\delta \approx |\omega_{g_1} - \omega_{g_2}|$), the force drops off significantly (Fig.~\ref{fig:intermediate}). Even adjusting the applied laser irradiance does not fully restore the force. In this regime, it is impossible to treat the second state either as an incoherent perturber or as a coherent contributor to the BCF.  Clearly this region of the parameter space should be avoided.

By contrast, as the detuning increases further and enters the large-detuning regime, the force re-emerges with an optimal $\Omega^{\textrm{tot}}$ lower than the optimal value for either transition individually.  This corresponds to the maximum visible in the lower right-hand corner of Fig.~\ref{fig:intermediate}.  In this high-detuning regime where $\delta >> |\omega_{g_1} - \omega_{g_2}|$, the splitting between the lower states becomes negligible and a transition is driven between the upper state and a superposition of the two lower states. The total Rabi frequency then approaches its two-level value of $\Omega^{\textrm{tot}} = \sqrt{3/2}\,\delta$.

\begin{figure}
\includegraphics[width=0.9\linewidth]{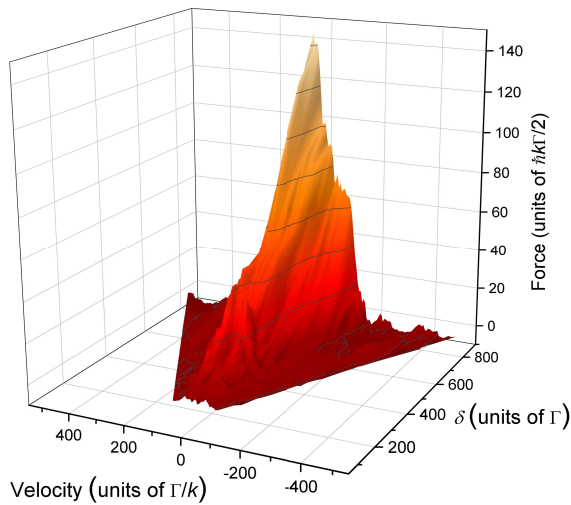}
\caption{(Color online) Force due to large-detuning BCF in the Fig.~\ref{fig:twoplusone} system. Data was smoothed over a velocity range of 0.1 $\delta/k$. Both the force near zero velocity and the velocity range of the force increase with increased bichromatic detuning.}
\label{fig:largedetuningforce}
\end{figure}

Once the dead zone of low force in the intermediate-detuning regime is exited, the force and velocity range both grow linearly with $\delta$, continuing up to quite large bichromatic detunings as shown in Fig.~\ref{fig:largedetuningforce}. The slopes for the force and the velocity range are found by least-squares fitting to be 0.258(22) $\hbar k/2$ and 0.123(21) $1/k$, respectively.

From these results, we can give a general prescription for large detunings that the optimal irradiance when multiple lower states are coupled to a single upper state is the one that produces a total quadrature-summed Rabi frequency of $\sqrt{3/2}\,\delta$ for the transitions ``surrounded" by the two bichromatic frequencies. In the small-detuning regime, where only one transition is addressed, the irradiance should instead be chosen so that Rabi frequency for this individual transition is $\sqrt{3/2}\,\delta$.  In both cases, we can also say that the force and its velocity range increase monotonically with $\delta$.  Finally, we have learned that the intermediate-detuning region generally leads to very small forces, and is to be avoided.

\section{Repumping}\label{sec:repumping}

For molecular applications of the BCF, it will often be the case that radiative decay losses to distant dark states are significant over the desired interaction time, requiring a repumping scheme to recover the population from these states.  To understand the effects of incoherent population loss and subsequent repumping in isolation from other complications, we have simulated a 1+1 system that also includes a distant dark state.  A continuous repumping laser field is used to transfer population from the dark state to the same excited state used by the cycling BCF transition. Because radiative decay is incoherent, the effects of multiple dark states are expected to be qualitatively similar to those simulated here for a single dark state, apart from the obvious impact of increased level degeneracies.

\begin{figure}
\includegraphics[width=0.9\linewidth]{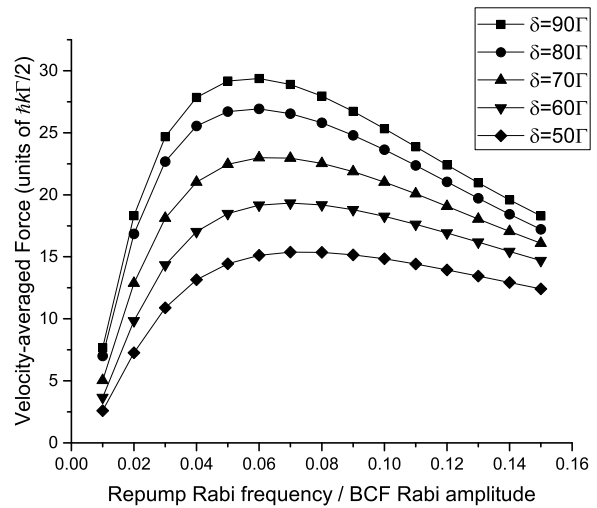}
\caption{In a 1+1 system with 5\% of decays going to a distant state, the achievable force depends on the Rabi frequency for repumping out of the distant state. The force exhibits a clear optimum in the repumping Rabi frequency that depends on the bichromatic detuning $\delta$. An optimum BCF Rabi amplitude of $\sqrt{3/2}\,\delta$ is assumed throughout. Solid lines are smooth curves to guide the eye.}
\label{fig:repump_omega}
\end{figure}

Figure~\ref{fig:repump_omega} shows the calculated bichromatic force as a function of the repumping laser Rabi frequency for an example in which which 5\% of decays are to the distant state.  For each of the several detunings $\delta$ shown in the figure, a broad but obvious maximum in the force is observed, indicating that there is an optimum value for the repumping laser irradiance.  This contrasts starkly with more familiar case of repumping for the incoherent radiative force, for which a simple saturation behavior is typically observed and the repumping irradiance is otherwise noncritical.  The difference can be attributed primarily to decoherence induced when the repumping rate is too rapid.

For the example in Fig.~\ref{fig:repump_omega}, the optimum repumping Rabi frequency was found to very closely fit a set of curves proportional to $\sqrt{\delta \Gamma}$, and this scaling is expected to be generally applicable.  Using a least-squares fit, the optimal frequency was determined to be
\begin{equation}
\Omega_{ed}^{\text{opt}} = 0.536(14) \sqrt{\delta \Gamma}.
\end{equation}

\begin{figure}
\includegraphics[width=1.0\linewidth]{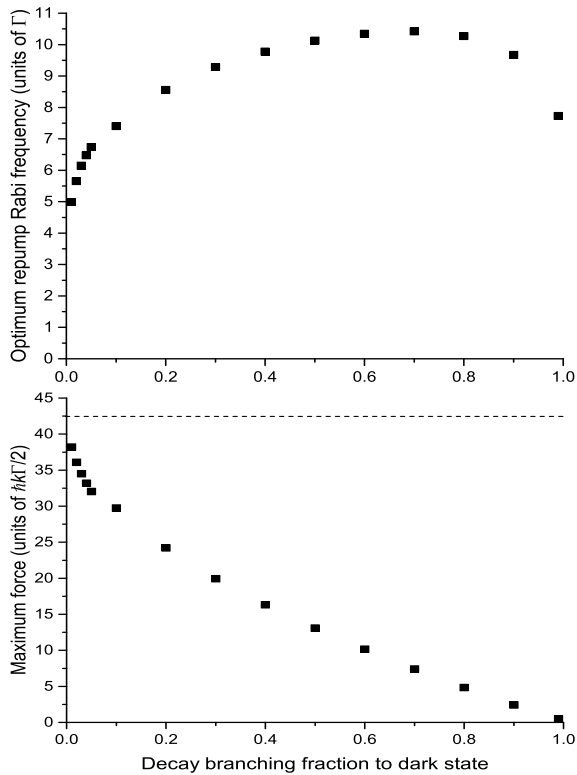}
\caption{At a fixed bichromatic detuning of $100\Gamma$, the optimal repump Rabi frequency (top) and the optimized force on molecules with near-zero velocities (bottom) depend on the branching ratio of excited state decays to a distant dark state. For very small decay rates to the dark state, the force approaches a value given by the two-level force multiplied by a statistical factor equal to the participating state fraction (dashed line, bottom).}
\label{fig:repump_maxes}
\end{figure}

Comparing the BCF with repumping at optimal parameters to the two-level BCF, the velocity range remains unchanged, but the peak force is reduced.  For the case of 5\% decay to a distant state, the force is reduced to approximately 50\% of the two-level force. Further simulations were carried out to compare the reduction in force over a range of decay branching ratios, with the bichromatic detuning and overall excited-state decay rate remaining fixed. As seen in Fig.~\ref{fig:repump_maxes}, as the branching fraction of spontaneous decays to the distant state increases, the peak force decreases monotonically. For very small dark-state branching fractions, the force converges to 2/3 of the two-level bichromatic force, the value expected based on steady-state population statistics. At large branching fractions, the peak force decreases approximately linearly to zero. The optimal repumping Rabi frequency increases with the decay fraction to the dark state over most of the range, but begins to decline when more than 70\% of the decay is to the dark state.

It should be kept in mind that one of the key advantages of the bichromatic force is the rapid stimulated transition rate, which allows a much larger total momentum transfer than incoherent forces if the available interaction time is limited.  Thus for weak out-of-system decay channels, it will often become possible to omit a repumping scheme and allow the loss, as the time it would take to accumulate a large population in the distant state may be long compared to the needed interaction time. Nevertheless, for the case of rapid decay to dark states the need for repumping will be unavoidable.

\section{Dark State Destabilization in Multilevel Systems}\label{sec:dark_destabilization}

Here we consider the effects of dark states within the transition manifold, which differ fundamentally from the distant incoherently-coupled dark states treated in Sec. \ref{sec:repumping}.  These dark states can form either when the electric dipole selection rules make certain levels inaccessible, or, for the case of degenerate levels, when a coherent dark state forms in which the excitation dipole amplitude to a superposition state coherently cancels.  For any system in which the ground-state multiplicity of degenerate projection quantum states $m$ exceeds the excited-state multiplicity by at least two, there will be two dark states regardless of the polarization of light used to couple the two manifolds \cite{Berkeland02}. A number of schemes have been developed to prevent or reverse optical pumping of the population into these dark states \cite{Berkeland02, Yeo15}, which would otherwise cause the BCF to rapidly diminish to zero.  We investigate the two most obvious schemes, one involving the application of a skewed dc magnetic field and the other, rapid switching of the optical polarization state.

\subsection{Skewed Magnetic Field}\label{subsec:skewmag}

The first method is to apply a dc magnetic field of magnitude $B$ at an angle $\theta_{BE}$ relative to the principal quantum axis defined by the optical polarization. For a state with angular momentum $J$ and degeneracy $2J+1$, this causes a remixing of the $m_J$ levels and leads to Zeeman shifts proportional (to first order) to $m_J$. This approach has been analyzed in detail for monochromatic excitation in Ref.~\cite{Berkeland02} and has recently been used successfully for radiative cooling and trapping of SrF by the DeMille group \cite{Barry14}. Here we define the magnetic field to lie in the $x$-$z$ plane, where it can be written in spherical tensor notation as

\begin{equation}
\vec{B} = \frac{B}{\sqrt{2}} \sin(\theta_{BE}) \hat{T}_1^{-1} + B \cos(\theta_{BE}) \hat{T}_1^0 - \frac{B}{\sqrt{2}} \sin(\theta_{BE}) \hat{T}_1^{+1}
\label{eq:sphericalmag}
\end{equation}

The magnetic-field Hamiltonian is modeled by adding the additional terms
\begin{eqnarray}
\label{eq:maghamiltonian}
\frac{H_B}{\hbar} &=& \mu_B \sum_{p,q}\bra{g_p}\vec{B}\cdot(g_s \hat{S}+ g_L \hat{L})\ket{g_q}\nonumber\\
&+& \mu_B \sum_{i,j}\bra{e_i}\vec{B}\cdot(g_s \hat{S}+ g_L \hat{L})\ket{e_j}
\end{eqnarray}
to the Hamiltonian given in Eq.~\ref{eq:hamiltonian}. Magnetic interactions with nuclear spins are neglected because they are generally small, and magnetic coupling between ground and excited states is neglected because for small dc magnetic fields this coupling is extremely small compared with typical optical energy level separations.

In Ref.~\cite{Berkeland02} the steady-state excited fraction is examined in a $^3S_1 \leftrightarrow {^3P}_0$ system driven by linearly-polarized monochromatic light with a skewed dc magnetic field. We have simulated this same system, replacing the monochromatic optical field with a BCF field. We scale the magnetic field units such that the first-order Zeeman shift is equal to $m_J B \cos(\theta_{BE})$, and express $B$ in units of the total upper-state decay rate $\Gamma$.

\begin{figure}
\includegraphics[width=0.9\linewidth]{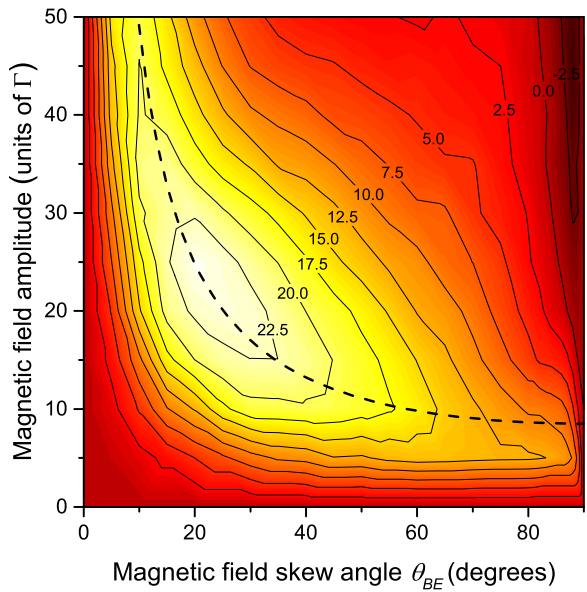}
\caption{(Color online) Contours of velocity-averaged force for a $^3S_1 \leftrightarrow {^3P}_0$ system driven by a linearly-polarized BCF field of $\delta=100\Gamma$ and $\Omega=122.5\Gamma$, with a dc magnetic field of varying amplitude and angle. The dashed line that follows the ridge of the force defined by $\sqrt{2} B \sin(\theta) = 12 \Gamma$.}
\label{fig:mag100}
\end{figure}

For a fixed bichromatic detuning $\delta=100\Gamma$ and Rabi frequency $\Omega^0=\sqrt{3/2}\,\delta$ for the single transition being directly driven, simulations with varying magnetic field strengths and skew angles reveal a clear optimum force in the $B$-$\theta_{BE}$ plane, evident in Fig.~\ref{fig:mag100}. There is also a ridge in the force that corresponds to a fixed remixing rate, up to a skew angle of about $60^\circ$. By fixing $\delta$, $B$, and $\theta_{BE}$ at their optimum values as determined by these simulations and varying $\Omega^0$, we confirm that a Rabi frequency of $\Omega^0 = \sqrt{3/2}\, \delta$ remains optimal in this system.

At varying detunings, always keeping $\Omega^0$ at its two-level optimum, the magnetic field parameter space was searched to find how the optimum parameters change with detuning. It was found that the optimal intra-ground-state coupling scales with the square root of the detuning and that the optimal level splitting scales linearly with the detuning, and we find using least-squares fitting that these optimal values are
\begin{eqnarray}
\sqrt{2}B \sin(\theta_{BE})& = &1.184(31) \sqrt{\delta \Gamma},\nonumber\\
B \cos(\theta_{BE})& = &0.2054(35) \delta.
\label{eq:magsincos}
\end{eqnarray}
The $\sqrt{2}B \sin(\theta_{BE})$ parameter plays the same role here as the repumping Rabi frequency in Sec.~\ref{sec:repumping}, which is to re-introduce lost population to the cycling transition. That both have optimum values that scale in the same way as $\sqrt{\delta \Gamma}$ stands to reason. Eq.~\ref{eq:magsincos} can be rearranged to give expressions for the optimum magnetic field strength and angle in this system, directly:

\begin{eqnarray}
B& = &\delta \sqrt{0.700(37) \Gamma/\delta + 0.0422(14)}\nonumber\\
\theta_{BE}& = &\tan^{-1}\left(4.07(13) \sqrt{\Gamma/\delta}\right)
\end{eqnarray}

\begin{figure}
\includegraphics[width=0.9\linewidth]{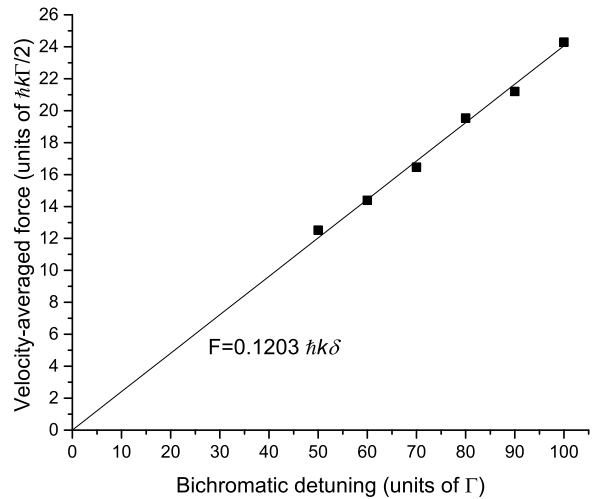}
\caption{At optimum magnetic field parameters and Rabi frequency, the bichromatic force with magnetic remixing in a $^3S_1 \leftrightarrow {^3P}_0$ system results in an average force that scales linearly with bichromatic detuning and is approximately one-third of the ideal two-level bichromatic force.}
\label{fig:magFvd}
\end{figure}

If the magnetic field parameters are kept optimized, the velocity-averaged bichromatic force remains linear with the bichromatic detuning as shown in Fig.~\ref{fig:magFvd}. A least-squares fits gives the constant of proportionality as $F/\delta = 0.1203(27) \hbar k$  . This is just slightly greater than 1/3 of the ideal two-level proportionality of $F/\delta = \hbar k/\pi$ ($1/3\pi = 0.1061$). Since the remixing process blends one participating ground state with two non-participating states, a multiplier close to 1/3 is indeed expected.

For other multilevel systems the details will differ, but the basics of this remixing scheme can be expected to be similar, as will the scaling rules.  An example of level mixing in a much more complicated system is given in Section \ref{subsec:cafsim}, where a detailed 16-level simulation is carried out for the bichromatic force on a molecular beam of CaF.

\subsection{Polarization Switching}

An alternative method for destabilizing dark states is to dynamically switch the polarization of the driving fields, thereby dynamically switching which state is dark. This can generally be accomplished using the counterpropagating-beam geometry of a BCF configuration, with the notable exception of four-level transitions for which $J''=1$ in the lower state and $J'=0$ in the upper state.  While the simple structure of these transition is often advantageous\cite{Gupta1993}, dark states can be avoided only if three polarization states are available, requiring a new laser beam at a skewed angle\cite{Berkeland02}.  This awkward arrangement strongly disfavors the polarization-switching method relative to the alternative of using a skewed $\textbf{B}$ field.

We consider transitions of the type $J' = J'' +1$ with $J''>1$.  Here the dark states can be avoided by alternating between $\sigma^+$ and $\sigma^-$ polarization. For example, in a $J''=2\leftrightarrow J'=1$ transition, defining the quantization axis along the $k$-vector of the stimulating light, $\sigma^+$ light leaves $m''=1$ and $m''=2$ dark, while $\sigma^-$ light leaves $m''=-1$ and $m''=-2$ dark. Clearly the dark-state spaces for these polarizations do not overlap, and the necessary polarization switching can readily be performed on a collinear pair of beams.

To test whether polarization switching during BCF still allows significant forces, a $^5 S_2 \leftrightarrow {}^5 P_1$ system was simulated with the polarization of the BCF field switched between $\sigma^+$ and  $\sigma^-$. Since the pulse area of the bichromatic beat notes is a critical BCF parameter, a sinusoidal variation between polarization states can be expected to give poor results because it leads to non-constant pulse areas. Instead, we simulate a simple on-off modulation pattern in which each polarization is kept on for a time $T$ before switching to the other polarization.

During each step in this cycle, population will gradually be optically pumped into the two dark $m_J$ states that cannot be re-excited, diminishing the force commensurately. Each switch in polarization constitutes a rephasing of the system after which coherence on the new transition (and thus the optical force) will gradually begin building up. An optimal switching period is therefore expected that balances these two effects.

The initial phase of the bichromatic optical field influences how quickly an incoherent population equilibrates under bichromatic driving. In most of our simulations, this phase is set to $\theta=45^\circ$, which gives the quickest equilibration, and the transient behavior is allowed to die away before evaluating the force. However, this transient behavior takes up a significant portion of each polarization-switching cycle and must now be treated more systematically. Simulations for which the global phase is reset to a particular value $\theta$ at each switch of polarization show that $\theta=45^\circ$ does indeed result in the strongest force, and $\theta=0^\circ$ the weakest, actually exhibiting a weak force in the reverse direction.

\begin{figure}
\includegraphics[width=0.9\linewidth]{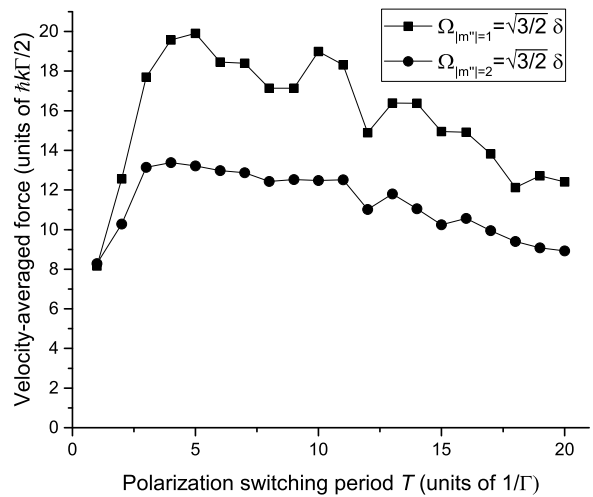}
\caption{Bichromatic force for a $^5S_2 \leftrightarrow {}^5P_1$ transition as a function of the switching period $T$ between $\sigma^+$ and $\sigma^-$ polarization states.  The Rabi frequency is set to its optimum value for the $|m''|=1\leftrightarrow m'=0$ transition or for the $|m''|=2\leftrightarrow |m'|=1$ transition, giving the two curves above, and the bichromatic detuning is $100\Gamma$ throughout.}
\label{fig:switching}
\end{figure}

The $^5 S_2 \leftrightarrow {}^5 P_1$ system has a further complication because the three transitions driven by $\sigma$-polarized light are not equally strong. Therefore it is not immediately apparent what laser irradiance is optimal. Simulating this polarization switching scheme at $\delta=100\Gamma$ over a range of switching times $T$ showed that a maximum in the force is achieved when setting the Rabi frequency to the value optimal for the $|m''|=1\leftrightarrow m'=0$ transition. Also, a switching time of $T\approx 5/\Gamma$ is found to be optimal as shown in Fig. ~\ref{fig:switching}, with a shallower fall-off on the long-$T$ side (Fig.) than for short $T$.. The peak force in these simulations is approximately $20 \hbar k \Gamma / 2$, which is about 1/3 of the peak two-level force at the same bichromatic detuning, $64 \hbar k \Gamma /2$.

If the switching cycle is not phase-locked to the global bichromatic phase, the random starting phase at each polarization switch will cause a reduction in the force compared with these simulations. As expected,the polarization switching period must be at least comparable to the upper state decay rate ($\sim\Gamma/5$ in these simulations) and the transition time between polarization states should be short compared to $1/\delta$, typically requiring nanosecond-scale switching.

For comparison, dark-state destabilization by a skewed magnetic field was also simulated for the same $^5 S_2 \leftrightarrow {}^5 P_1$ model system, with $\pi$-polarized BCF fields. With the Rabi frequency optimized for $|m=1|$ transitions, such that two of the three available transitions were being driven with optimal Rabi frequency, the peak force over a considerable range of magnetic field parameters is found to be $24 \hbar k \Gamma /2$.  Unlike the polarization switching scheme, no fine-tuning of phases or timing is required.  At least in this system, magnetic field remixing is the better choice, but if a particular application cannot tolerate external magnetic fields, polarization switching remains a viable alternative.

\section{Calcium Monofluoride}\label{sec:caf}

\subsection{Structure and Parameters}\label{subsec:structure}
\begin{figure}
\includegraphics[width=0.9\linewidth]{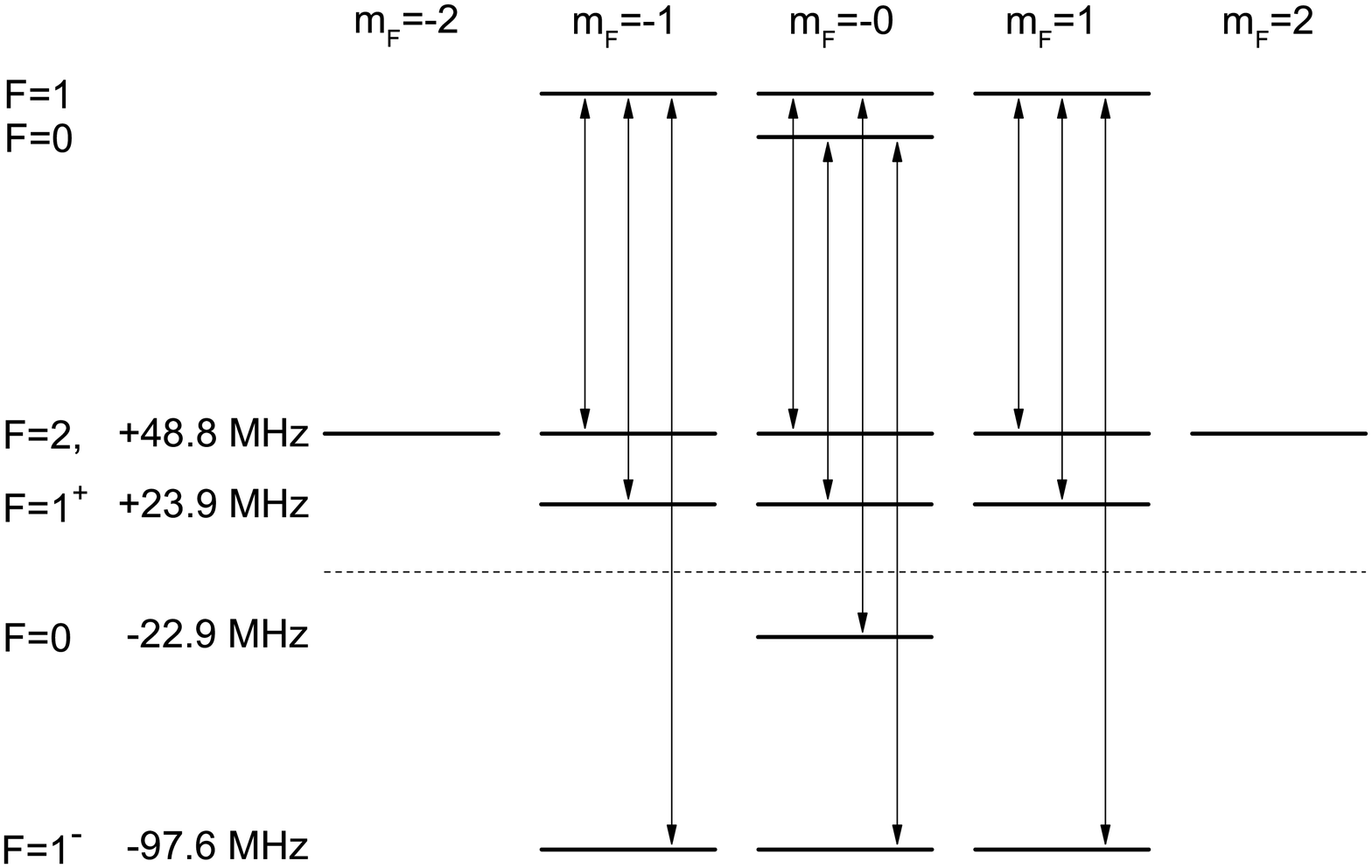}
\caption{The $B \leftrightarrow X$ $P_{11}(1.5)/^P Q_{12}(0.5)$ system in CaF, showing zero-field energy spacings. The hyperfine splitting of the $B$ levels is taken to be negligible compared to the $X$ state. The allowed transitions for $\pi$-polarized light are indicated.}
\label{fig:leveldiagram}
\end{figure}

With an eye towards experimental realization of the BCF using calcium monofluoride, the structure of the $^2\Sigma^{+}B \leftrightarrow {}^2\Sigma^{+}X$ $(0-0)$ $P_{11}(1.5)/^P Q_{12}(0.5)$ branch was examined. This scheme differs in some details from the $A \leftrightarrow X$ transition that we have previously discussed semi-quantitatively in Ref.~\cite{Chieda11}, although the differences are for the most part minor.  The fine and hyperfine sublevels for this transition are shown in Fig. \ref{fig:leveldiagram}.  In the $X$ state $N$=1 manifold, $J$ is a poor quantum number due to mixing of the two $F$=$N$=1 levels. The two eigenfunctions $\ket{F=1^\pm}$ can be written in terms of the $J$=1/2 and $J$=3/2 levels with $F$=1 by introducing a mixing angle $\phi$,
\begin{equation}
\begin{pmatrix}
  \ket{F\textrm{=1}^+,m_F} \vspace{3pt}\\
  \ket{F\textrm{=1}^-,m_F}
\end{pmatrix}
=
\begin{pmatrix}
  \cos\phi & \sin\phi \vspace{3pt} \\
  -\sin\phi & \cos\phi
\end{pmatrix}
\begin{pmatrix}
\ket{3/2,m_F}\vspace{3pt} \\
\ket{1/2,m_F}
\end{pmatrix}.
\label{eq:jmixing}
\end{equation}
The mixing angle was determined to be $41.18^\circ$ via the methods of Ref.~\cite{Sauer1996}, in which the structurally similar YbF molecule was examined.  We used the molecular constants for CaF listed in Ref.~\cite{Childs1981}.

When the pure $J$ states are expressed in case ($a_\beta$) notation, $\ket{\Lambda,S,\Sigma,\Omega,J,I,F,M_F}$, the electric dipole transition matrix elements can be calculated as in Ref. \cite{Wall2008} up to a shared factor $\bra{\Lambda', S, \Sigma' = 1/2} T_0^{(1)} (\hat{d}) \ket{\Lambda, S, \Sigma = 1/2}$. Additionally, the squared sum of matrix elements is normalized to this factor,
\begin{eqnarray}
\sum_{p} \sum_{J'',F'',M''_F}&&\left| \bra{J',F',M_F'}T_p^{(1)} (\hat{d}) \ket{J'',F'',M_F''} \right|^2 \nonumber \\& = &\left| \bra{\Lambda', S, 1/2} T_0^{(1)} (\hat{d}) \ket{\Lambda, S, 1/2} \right|^2.
\end{eqnarray}

We define a relative dipole transition element between states $\ket{i}$ and $\ket{j}$ as
\begin{equation}
\label{eq:relativedipole}
\kappa_{ij} = \frac{\sum_p \bra{i} T_p^{(1)} (\hat{d}) \ket{j}}{\bra{\Lambda', S, \Sigma' = 1/2} T_0^{(1)} (\hat{d}) \ket{\Lambda, S, \Sigma = 1/2}}.
\end{equation}
We calculate these relative dipole transition matrix elements, following Ref.~\cite{Wall2008} with the exception that $\Delta\Lambda = 0$ rather than $\pm 1$, and then introduce the $J$-mixing discussed above to obtain the relative dipole transition matrix elements between the eigenfunctions (Table~\ref{tab:btox}).

\begin{table}
\caption{\label{tab:btox}
Electric dipole matrix elements for the $B \leftrightarrow X$ $P_{11}(0.5) / ^PQ_{12}(1.5)$ branch in CaF, in units of the total $B\leftrightarrow X$ dipole transition moment.}
\begin{ruledtabular}
 \begin{tabular}{lr|d r@{ } ddd}
  & \multicolumn{1}{r}{}&
  \multicolumn{1}{c} {$F'$=0} &&
  \multicolumn{3}{c} {$F'$=1}\\
 \cline{3-4} \cline{5-7} \\
 $F''$ & $m_{F}''$ &
 \multicolumn{1}{c} {$m_F'$=0} &&
 \multicolumn{1}{c} {$m_F'$=$-1$} &
 \multicolumn{1}{c} {$m_F'$=0} &
 \multicolumn{1}{c} {$m_F'$=1} \\
 \hline
 & $-2$ & 0 && -0.5774 & 0 & 0 \\
 & $-1$ & 0 && 0.4082 & -0.4082 & 0 \\
2 & 0 & 0 && -0.2357 & 0.4714 & -0.2357 \\
 & 1 & 0&&  0 & -0.4082 & 0.4082 \\
 & 2 & 0 && 0 & 0 & -0.5774 \\
 &&& \\
 & $-1$ & -0.5743 && -0.0421 & -0.0421 & 0 \\
$1^+$ & 0 & 0.5743 && 0.0421 & 0 & -0.0421 \\
 & 1 & -0.5743 && 0 & 0.0421 & 0.0421 \\
 &&& \\
 & $-1$ & 0.0595 && -0.4061 & -0.4061 & 0 \\
$1^-$ & 0 & -0.0595 && 0.4061 & 0 & -0.4061 \\
 & 1 & 0.0595 && 0 & 0.4061 & 0.4061 \\
 &&& \\
0 & 0 & 0 && 0.3333 & 0.3333 & 0.3333 \\
\end{tabular}
\end{ruledtabular}
\end{table}

The normalization factor can be estimated by working back from the measured radiative decay lifetime of the $B$ state, $\tau_B = 25.1 \pm 4$~ns \cite{Zare1973}. The characteristic lifetime of a state before it decays along a particular transition is given by the inverse of the Einstein A coefficient,
\begin{equation}
\tau_{ij} = 1/A_{ij} = \frac{3 h \lambda^3 \varepsilon_0}{16\pi^3 \left| d_{ij} \right|^2}.
\label{eq:channellifetime}
\end{equation}
The total lifetime of a state is the inverse of the sum of the $A$ coefficients for all possible decays. Outside of the channels we are considering, other rotational decays on $B \leftrightarrow X$ are forbidden, but other vibrational decays are allowed, as are decays to other electronic states. The Franck-Condon factor for the $B \leftrightarrow X$ (0-0) band has been calculated to be 0.999 \cite{Dulick1980}. The decay branching ratios to different electronic states scale as $\left| d_{ij}\right|^2 \nu_{ij}^3$. The dipole transition moments have been estimated via ligand field modeling \cite{Rice1985} and the transition frequencies are known \cite{Huber1973}, so the ratio of decay rates to the can be estimated:
\begin{eqnarray}
\label{electronicratios}
\Gamma_{B \rightarrow X}& : & \Gamma_{B\rightarrow A} \nonumber\\
= (1.73 \text{ ea}_0)^2 (18833\,\text{cm}^{-1})^3& : &(0.58 \text{ ea}_0)^2 (2307\,\text{cm}^{-1})^3\nonumber\\
= 4840:1.
\end{eqnarray}
Thus, approximately 99.98\% of radiative decays from the $B$ state produce molecules in the $X$ state. Multiplying the electronic, vibrational and rotational factors, 99.8\% of decays from $B$, $v$=0, $N$=0 produce the $X$, $v$=0, $N$=1 state.

This allows an estimate of the total $A$ coefficient for the $B \leftrightarrow X$ $P_{11}(0.5) / ^PQ_{12}(1.5)$ branch,
\begin{equation} \label{eq:dipoleeq}
A = \frac{0.998}{\tau_B} = \frac{16 \pi^3 \left| \bra{\Lambda', S, 1/2} T_0^{(1)} (\hat{d}) \ket{\Lambda, S, 1/2} \right|^2}{3 h \varepsilon_0 \lambda},
\end{equation}
where $\lambda = 530.9 \pm 0.1$ nm is the transition wavelength.  Rearranging, we can estimate that the matrix element is
\begin{equation}
\label{eq:dipolevalue}
\left| \bra{\Lambda', S, 1/2} T_0^{(1)} (\hat{d}) \ket{\Lambda, S, 1/2} \right| = 1.7(1) e a_0,
\end{equation}
which is consistent with the value from Ref.~\cite{Rice1985} of 1.73 ea$_0$. Combining Eqs.~\ref{eq:rabiamp} and \ref{eq:relativedipole} with the irradiance of a monochromatic plane wave gives practical units for the Rabi frequency amplitude of a transition in this channel for a given per-bichromatic-component irradiance $I$ expressed in W/cm$^2$,
\begin{equation}
\label{eq:cafrabi}
\Omega_{ij}^0 = \kappa_{ij} \sqrt{I}\, (2\pi \times 60\mbox{ MHz}),
\end{equation}
assuming that the light is optimally polarized to drive the transition.

 Assuming that magnetic remixing of dark states is used, the response to a dc magnetic field must be understood as well. The magnetic Hamiltonian in Eq.~\ref{eq:maghamiltonian} can be evaluated using equations (8.183) and (8.184) of Ref.~\cite{Brown2003}, with a few obvious modifications to allow for the spherical components $p \neq 0$. These equations give the complete picture, but the six $F$ states can be generally characterized in a more intuitive way via their $g$-factors $g_F$, which for the two $F''=1$ levels must take the $J$-mixing of the eigenstates into account (Table~\ref{tab:gfactors}).

\begin{table}
\caption{\label{tab:gfactors}
 $g$-factors for the six $F$ levels involved in the $B \leftrightarrow X$ $P_{11}(0.5) / ^PQ_{12}(1.5)$ branch in CaF, scaled by the electron spin $g$-factor $g_s$.}
\begin{ruledtabular}
\begin{tabular}{r | c c c c c c c}
&$F''$=0&$F''$=1$^-$&$F''$=1$^+$&$F''$=2&$F'$=0&$F'$=1\\
\colrule
$g_F/g_s$&0.0&0.320&-0.070&0.25&0.0&0.5
\end{tabular}
\end{ruledtabular}
\end{table}

\subsection{Simulations}\label{subsec:cafsim}

Previous treatments of the bichromatic force in multilevel systems \cite{Chieda11,Dai15} have been restricted to analyses based on two-level calculations combined with statistical arguments. These treatments cannot account for multilevel coherent effects such as those seen in Sec.~\ref{sec:twoplusone}, which were included along with detuning and line-strength considerations in a poorly-determined ``force reduction factor," $\eta$. Examining the structure of the $^2\Sigma^{+}B \leftrightarrow {}^2\Sigma^{+}X$ $(0-0)$ $P_{11}(1.5)/^P Q_{12}(0.5)$ transition in CaF (Fig. \ref{fig:leveldiagram}), it is easy to see that it cannot be fully decomposed into two-level systems even for the simplest case of $\pi$-polarized light.  Each excited-state sublevel is coupled to either two or three ground-state levels, and thus the coherent effect contribution to $\eta$ cannot be neglected.

However, Fig. \ref{fig:leveldiagram} also suggests that some simplification may be possible without giving up much accuracy.  No ground-state sublevel is laser-coupled to more than one excited-state sublevel, so that the entire system can be viewed as two 2+1 and two 3+1 subsystems which are weakly coupled to each other by incoherent radiative decays, in addition to two dark ground-state levels. We start by taking this simplified approach and modeling these four subsystems independently, neglecting the radiative coupling between them. Because the coherence terms of the density matrix decay at a rate set by the total excited-state decay rates, we proportionally increase the decay rates within each subsystem to compensate for the neglected decays.

\begin{figure}
\includegraphics[width=\linewidth]{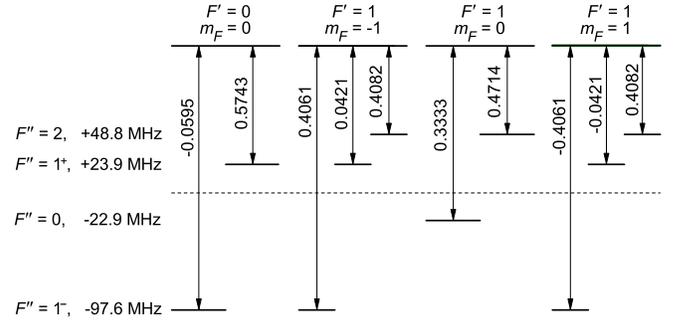}
\caption{The four subsystems defined by isolating the $\Delta m_F = 0$ transitions of the $B \leftrightarrow X$ $P_{11}(1.5)/^P Q_{12}(0.5)$ branches in CaF. Relative dipole transition matrix elements are indicated for each transition.}
\label{fig:subsystems}
\end{figure}

The levels in the $X, N''=1$ states span 146.4 MHz, or 23.1 $\Gamma$ in units of the excited-state decay rate. As seen in Sec.~\ref{sec:twoplusone}, so long as the bichromatic detuning is larger than the energy splitting, the optimal $\Omega^{tot}$ will be $\sqrt{3/2}\,\delta$. Within each of the four subsystems, the quadrature sum of the $\kappa_{ij}$ is $1/\sqrt{3}$, so the same laser irradiance can drive every one of the subsystems with the same $\Omega^{tot}$.

\begin{figure}
\includegraphics[width=0.9\linewidth]{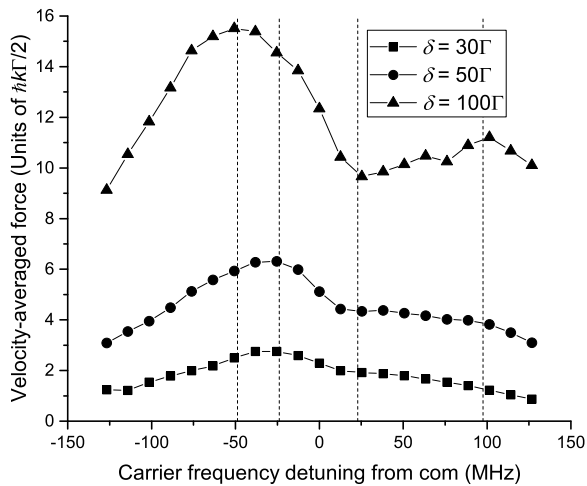}
\caption{A weighted average of the force on isolated subsystems of the $B \leftrightarrow X$ $P_{11}(1.5)/^P Q_{12}(0.5)$ branches in CaF gives an estimate for the optimal carrier frequency detuning from the transition center of mass of the system and for the peak force. Resonances are indicated with vertical dashed lines.}
\label{fig:subsystemaiming}
\end{figure}

Using the optimal $\Omega^{tot}$ at each of $\delta=30\Gamma$, $50\Gamma$ and $100\Gamma$, each of the four subsystems was independently simulated for a range of BCF carrier frequencies that spanned the entire set of transitions. The results from each subsystem were then combined together with a weighting given by the fraction of ground state sublevels included in each subsystem. This weighting is based on the assumption that the system begins with a statistical distribution of population in the ground states and that each subsystem will maintain that population throughout the interaction with the BCF fields. As shown in Fig.~\ref{fig:subsystemaiming}, the averaged force on molecules with near-zero velocities was found to peak at a carrier frequency detuned somewhat from the hyperfine center of mass, at a location between the $F''=2$ and $F''=1^+$ components. Transitions from these two levels were the strongest transitions in each of the four subsystems.

Using optimal parameter values, estimated values for the key properties relevant to longitudinal slowing of a molecular beam are summarized in Table~\ref{tab:cafsummary} in the columns labeled as Method (b).  The table also includes estimates of the these same quantities based on the semi-quantitative statistical method of Ref.~\cite{Chieda11}, where applicable, as well as estimates based on the complete 16-level simulation that we describe next.  The subsystem approach must be regarded as an approximation both because it relies on assumptions about the population balance, and because it cannot include the effects of remixing dark states.  Instead it is assumed that on aggregate, a good remixing scheme will keep the dark populations at about the statistical fraction determined by the total number of states. The results of Sec.~\ref{sec:dark_destabilization} support this estimate, but the details are revealed only by modeling the full system.
\begin{table*}[ht]
\caption{\label{tab:cafsummary}
A summary of results for three different methods of estimating the BCF for $\pi$-polarized light in the sixteen-level $B \leftrightarrow X$ $P_{11}(0.5) / ^PQ_{12}(1.5)$ system in CaF, assuming no repumping. The methods are (a) the statistics-based estimate described in Ref.~\cite{Chieda11}, (b) a weighted subsystem-based multilevel simulation, and (c) a full 16-level simulation of the complete system, including dark-state remixing by a dc magnetic field. Some results for Method (a) includes a ``force reduction factor" $\eta$ as in Ref.~\cite{Chieda11}, discussed further in the text.  In the final row of the table, we consider the fraction of molecules remaining after slowing by 60 m/s, taking into account decays into distant dark states.}
\begin{ruledtabular}
\begin{tabular}{r | r@{\hspace{9pt}}r@{\hspace{9pt}}r@{\hspace{9pt}}|r@{\hspace{9pt}}r@{\hspace{9pt}}r@{\hspace{9pt}}|r@{\hspace{9pt}}r@{\hspace{9pt}}r@{\hspace{9pt}}}
& \multicolumn{3}{c}{$\delta$=30$\Gamma$}&
\multicolumn{3}{c}{$\delta$=50$\Gamma$}&
\multicolumn{3}{c}{$\delta$=100$\Gamma$}\\
&\multicolumn{1}{c}{(a)}&\multicolumn{1}{c}{(b)}&\multicolumn{1}{c}{(c)}&\multicolumn{1}{c}{(a)}&\multicolumn{1}{c}{(b)}&\multicolumn{1}{c}{(c)}&\multicolumn{1}{c}{(a)}&\multicolumn{1}{c}{(b)}&\multicolumn{1}{c}{(c)}\\
\colrule
Irradiance (W/cm$^2$)&45.2&45.2&45.2&125.6&125.6&125.6&503&503&503\\
Carrier detuning from c.o.m. (MHz)&N/A&$-32$&$-38$&N/A&$-30$&$-38$&N/A&-47&$-38$\\
Magnetic field magnitude (Gauss)&---&---&29.2&---&---&29.4&---&---&37.1\\
Magnetic field angle&---&---&71$^\circ$&---&---&61$^\circ$&---&---&66$^\circ$\\
Force near zero velocity ($\hbar k \Gamma/2$)&16.4$\eta$&2.8&2.6&27.3$\eta$&6.3&5.8&54.6$\eta$&15.5&14.2\\
Velocity range (m/s)&100&110&120&170&130&170&340&190&250\\
Average excited fraction&0.14&0.23&0.18&0.14&0.21&0.15&0.14&0.18&0.14\\
Time to slow by 60 m/s ($\mu$s)&14/$\eta$&84&91&8.7/$\eta$&38&41&4.3/$\eta$&15&17\\
Fully-slowed population fraction&---&0.21&0.27&---&0.53&0.61&---&0.80&0.83
\end{tabular}
\end{ruledtabular}
\end{table*}

\begin{figure}
\includegraphics[width=0.9\linewidth]{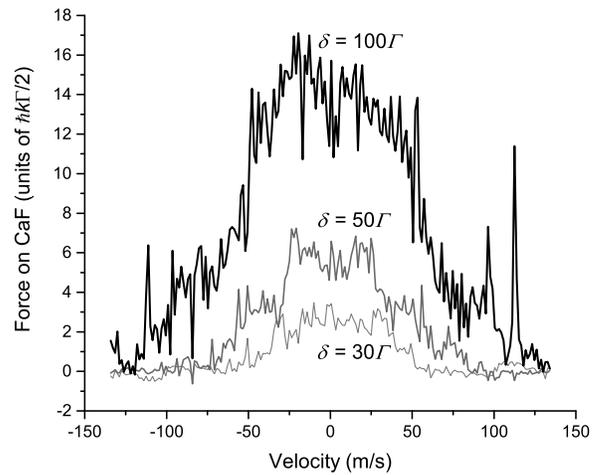}
\caption{Simulations of the bichromatic force for the sixteen-level $B \leftrightarrow X$ $P_{11}(0.5) / ^PQ_{12}(1.5)$ system in CaF, with dark-state remixing by a dc magnetic field. These forces far exceed the radiative force of $\sim$0.024~$\hbar k \Gamma/2$ realized in Ref. \cite{Zhelyazkova14}. The strength and velocity range of the force increase with the bichromatic detuning in the range of detunings simulated.}
\label{fig:fullcaf}
\end{figure}

Using the optimal laser irradiance determined from subsystem-based modeling at each bichromatic detuning, the entire set of sixteen levels was then simulated for $\pi$-polarized BCF fields and a skewed dc magnetic mixing field, using a two-step process. First, the remaining parameter space consisting of the magnitude and angle of the magnetic field and the carrier frequency detuning was surveyed at each of several small molecular velocities, in order to find optimal parameters. Then, using those parameters, simulations across the full velocity range were carried out.  The resultant force profiles are shown in Fig.~\ref{fig:fullcaf} for each of three representative bichromatic detunings, and the properties relevant to longitudinal slowing of a molecular beam are listed in the columns of Table~\ref{tab:cafsummary} labeled as Method (c).

As seen previously for 2+1 systems in Sec. \ref{sec:twoplusone} for detunings in the lower portion of the large-detuning regime, the force increases slightly faster than linearly with $\delta$.  Comparing to the results from the subsystem approach, it is clear that the subsystem approach consistently overestimates the peak force and the average excited state fraction, and it underestimates the velocity range of the force. However, all of these errors are small, with less than 10\% error in the force and less than 25\% error in the velocity range across all three simulated detunings.  In this example the subsystem-based calculations are not only simpler, but they also required only 1/5 of the computational time needed for the full sixteen-level calculation.  These considerations may make the simpler and more approximate subsystem method preferable for calculations intended for planning purposes.

Compared to both of our multilevel models, the statistical approach of Ref. \cite{Chieda11} is, of course, much less accurate.  It slightly underestimates the velocity range of the force at $\delta=30\Gamma$ but this crosses over to a large overestimate at $\delta=100\Gamma$. This can be attributed to the approximation that $\Delta v=\delta/k$ that does not accont for destructive coherent effects that limit the range of the force in the intermediate detuning regime. There is also a substantial overall uncertainty due to the ``force reduction factor" $\eta$, which essentially lumps together all of the effects not included in the effective two-level model.  Rough estimates in the spirit of Ref. \cite{Chieda11} predict that $\eta$ should be about 0.57 at $\delta$ = 30$\Gamma$, increasing to about 0.8 at 100$\Gamma$.  By comparing the estimated force to our full 16-level simulations, we can empirically evaluate the actual value of $\eta$ in each case, giving $\eta$ = 0.16, 0.21, and 0.26 for $\delta$ = 30$\Gamma$, 50$\Gamma$, and 100$\Gamma$ respectively. This verifies that the statistical approach is useful for making rapid order-of-magnitude estimates, and that as expected the value of $\eta$ gradually increases with detuning.  However, it also verifies that the actual numbers are not reliable beyond this level of approximation.  The method is useful for determining feasibility, but is not necessarily reliable for determining detailed experimental designs.

We conclude by comparing our the multilevel BCF with the radiative force, working from our sixteen-level simulation results.  Even for the lowest detuning simulated, the calculated bichromatic force is already much greater than the largest radiative force achieved in CaF to date. Using the same $X$-state level in an $A \leftrightarrow X$ rotationally-closed transition, a radiative force with a peak of 0.024(6) $\hbar k \Gamma/2$ has been realized \cite{Zhelyazkova14}. Our result of 2.6 $\hbar k \Gamma/2$ at $\delta=30\Gamma$ is larger by more than two orders of magnitude. This implicitly assumes that the BCF scheme does not require the $v_{10}$ repumping laser used in Ref. \cite{Zhelyazkova14}. Because the decay branching fraction to dark states is low, the results of Sec.~\ref{sec:repumping} indicate that even with repumping, the force would be reduced only by a statistical factor of at most a single order of magnitude.  However, we now take a closer look to demonstrate that the BCF acts rapidly enough to make repumping unnecessary for a practical decelerator that operates on a buffer-gas-cooled molecular beam.

In Sec.~\ref{subsec:structure}, we estimate that 0.2\% of radiative decays populate non-cycling states. The characteristic out-of-system decay time is this $\tau_B / (0.002 P_e)$, where $P_e$ is the average excited-state fraction. We find $P_e$ by averaging the calculated excited-state fraction over the velocity range in which the force is large.  The result is $P_e \approx 1/6$, with variations of just a few percent between the three simulations in Fig. \ref{fig:fullcaf}.  This value is also quite close to the estimate of 1/7 provided by the statistical approach of Ref.~\cite{Chieda11}. Thus we can estimate an out-of-system decay time of about 75 $\mu$s.

In recent work elsewhere, radiative slowing was used to slow a cryogenic beam of SrF by 60 m/s, and this proved sufficient to load a molecular MOT \cite{Barry12}.  A reasonable objective for CaF is to slow it by a comparable amount.  Using the bichromatic force with a detuning of $\delta=30\Gamma$, we estimate that 27\% of the illuminated CaF molecules with $N''$=1 would be slowed by 60 m/s without repumping, and this percentage increases to 83\% for $\delta=100\Gamma$. These results clearly indicate that as long as the required laser irradiance is available to drive the BCF with detunings of at least 30$\Gamma$, a repump-free optical decelerator is feasible for CaF and other molecules with similar near-cycling transitions.

\section{Conclusions}

If the coherent optical bichromatic force is to be applied to molecules or complex atoms, it is essential to prove that it can work for an intricately coupled multilevel system, and to have a reliable method for modeling such systems so that an optimal BCF scheme can be devised.  We have demonstrated that direct numerical integration of the density matrix provides stable and useful solutions for systems with up to 16 sublevels, using only the computational capacity of an ordinary personal computer.

For three-level lambda-type systems, we have demonstrated and quantitatively modeled a nonlinear dependence of the bichromatic force on detuning at small detunings. We have also shown that intra-ground-state coherences play a vital role at large detunings, allowing lower laser power requirements than would be estimated from individual transitions, with the general prescription that the quadrature sum of the Rabi frequencies of the addressed transitions should be equal to $\sqrt{3/2}\,\delta$. Further, we have shown that the region of parameter space where the bichromatic detuning is roughly equal to the transition energy splittings is to be avoided, as it results in severe reductions in the achievable force.

When an additional distant dark state is introduced, we show that a repumping laser can be effective, although the requirements for the laser power are more specific than for repumping with the ordinary incoherent radiative force.  For transitions involving more than three sublevels, there are additional complications due to the formation of dark states \textit{within} the system, but we show that this can be addressed by adding a small skewed magnetic field or by rapidly switching the optical polarization state. We have found that the magnetic field method is generally preferable.

We have shown that for many-level systems, a simulation approach in which the system is decomposed into individually-evaluated coherent subsystems provides significant computational speed-up with associated errors of less than 10\% in the predicted force.

Most importantly, we have quantitatively verified the predictions of our 2011 paper, clearly demonstrating that BCF slowing and cooling of molecules should be both practical and promising.  Examining the specific example of CaF, we show that the required laser power is readily attainable and the magnitude of the force is over two orders of magnitude larger than the radiative force for the same system.  Further, the momentum transfer prior to out-of-system decay is large enough to allow complete deceleration of a cryogenic beam of CaF without the use of repumping lasers and over a distance scale of just a single centimeter.

Similar gains relative to the radiative force can be expected for other small molecules.  In general, limited momentum transfer sufficient to produce state-selective transverse deflection should be feasible for almost any small molecule.  By contrast, substantial longitudinal slowing requires a near-cycling transition because the required momentum transfer involves thousands of photons, but the requirements are still much relaxed compared with the conventional radiative force.

\section{Acknowledgments}

This work was supported by the National Science Foundation and the University of Connecticut.


\begin{thebibliography}{}
\bibitem{Mathematica}
Versions 8 through 10 of Mathematica have been used during this work, with no significant difference noticed between versions.  The best convergence is generally obtained using Mathematica's standard ``ExplicitRungeKutta" method. Only minor improvements in speed were found using other computational environments, including a Fortran program for the specific case of a two-level system.
\end{thebibliography}
\end{document}